\documentclass[12pt]{article}
\usepackage{amsmath,amssymb}
\usepackage[dvips]{graphicx}
\usepackage{cite}

\newcommand{\bbibitem}[1]{\bibitem{#1}\marginpar{#1}}

\def\Label#1{\label{#1}%
  \smash{\hbox to0pt{\raise1ex\hbox{\tiny[#1]}\hss}}}
\def\noLabels{\let\Label=\label}
\def\nobbibitem{\let\bbibitem=\bibitem}

\newcommand{\be}{\begin{equation}}
\newcommand{\ee}{\end{equation}}
\newcommand{\bea}{\begin{eqnarray}}
\newcommand{\eea}{\end{eqnarray}}

\newcommand{\beq} {\begin{equation}}
\newcommand{\eeq} {\end{equation}}
\newcommand{\beqa} {\begin{eqnarray}}
\newcommand{\eeqa} {\end{eqnarray}}

\newcommand{\nn}{\nonumber}
\newcommand{\eq}[1]{(\ref{#1})}

\newcommand{\pat}{\partial}

\def\vek{\vec{k}}

\begin{document}

\rightline{HIP-2008-43/TH} \vskip 2cm \centerline{\Large {\bf The
energy-momentum tensor of perturbed tachyon matter}} \vskip 1cm
\renewcommand{\thefootnote}{\fnsymbol{footnote}}
\centerline{{\bf Niko
Jokela,$^{1,2}$\footnote{najokela@physics.technion.ac.il} Matti
J\"arvinen,$^{3}$\footnote{mjarvine@ifk.sdu.dk} and Esko
Keski-Vakkuri$^{4,5}$\footnote{esko.keski-vakkuri@helsinki.fi}}}
\vskip .5cm \centerline{\it ${}^{1}$ Department of Physics}
\centerline{\it Technion, Haifa 3200, Israel}
\centerline{\it
${}^{2}$ Department of Mathematics and Physics}
\centerline{\it University of Haifa at Oranim}
\centerline{\it Tivon 36006, Israel}
\centerline{\it ${}^{3}$Center for High Energy Physics, University of Southern
Denmark}
\centerline{\it Campusvej 55, DK-5230 Odense M, Denmark}
\centerline{\it
${}^{4}$Helsinki Institute of Physics and ${}^{5}$Department of
Physics } \centerline{\it P.O.Box 64, FIN-00014 University of
Helsinki, Finland}

\setcounter{footnote}{0}
\renewcommand{\thefootnote}{\arabic{footnote}}

\begin{abstract}
We add an initial nonhomogeneous perturbation to an otherwise
homogeneous condensing tachyon background and compute its space time
energy-momentum tensor from worldsheet string theory. We show that
in the far future the energy-momentum tensor corresponds to
nonhomogeneous pressureless tachyon matter.
\end{abstract}

\newpage

\paragraph{Introduction}
The condensation of the tachyonic mode (or modes) present in an
unstable brane system in string theory was found to lead to
pressureless tachyon matter \cite{Sen:2002nu,Sen:2002in,Sen:2002an}. This process has generated wide
interest in cosmology. Initially, it was speculated that tachyon matter
could be a novel form of dark matter, but since then tachyon condensation has
been applied to a variety of different scenarios, too numerous to attempt to
give a fair list of references. In this short paper we
will focus on one aspect. From very early
on, there has been interest in understanding the effect of fluctuations
to the condensing tachyon field, and their cosmological
implications. This was first studied in the simple effective tachyon
field theory model in \cite{Frolov:2002rr}. It is important -- but also more challenging --
to go beyond the simple toy model and to understand how initial
fluctuations could be taken into account and analyzed in string
theory. A first elementary question could be to see what kind of an
imprint small fluctuations leave on the energy-momentum tensor of
the resulting tachyon matter.  Recent new tricks, applying random matrix theory
to worldsheet calculations in a condensing tachyon background, have made it
possible to study the problem analytically in this paper. The answer should be of interest, 
{\em e.g.}, to cosmological models where gravity waves are generated by tachyon inhomogeneities 
\cite{Mazumdar:2008up}.

We focus on bosonic open string theory. In the presence of a background open string tachyon field,
the worldsheet action is
\beq
S_{\rm } = S_0 + S_T = \frac{1}{2\pi}\int_{\rm disk} (-\pat X^0\bar\pat X^0 + \pat X^I\bar\pat X^I)
 + \lambda\oint dt~T(X(t)) \ .
\eeq
Suppose that $T(X)$ is divided into a homogeneous rolling
background (taken to be the simple exponential profile) and small nonhomogeneous
perturbations,
\beq
 T(X(t)) = \lambda e^{X^0(t)} + \delta T(X^0,\vec{X}) \ .
\eeq
As is well known, the possible field configurations must satisfy the beta function equations
which can be interpreted as the equations of motion of the effective field theory.
Let us consider what this means for the perturbations $\delta T$, and turn
off the homogeneous profile for a moment.

The perturbations should be marginal deformations,
{\em i.e.}, in the expansion
\beq\label{pert1}
  \delta T(X^0,\vec X) = \int d\vek~\delta \hat T_{\vek}\int dt~ e^{-i\omega_k X^0(t)+i \vek \cdot \vec{X}(t)} \ ,
\eeq
the frequency $\omega_k$ is not independent but must satisfy the tachyon on-shell condition
\beq\label{onshell}
\omega^2_k-\vek^2=-1 \ .
\eeq
However, since the deformation is in the exponent inside the worldsheet path integral,
for a sizable deviation the deformation must continue
to be marginal even with respect to an already deformed theory. This means that the deforming operator must
be exactly marginal (or ``mutually self-local'', see \cite{Recknagel:1998ih});
for the tachyon operator
this condition can be satisfied only if $\omega=\pm i$ (leading to the homogeneous deformation) or
$\omega = \pm i/\sqrt{2},|\vek|=1/\sqrt{2}$. The latter condition means that a nonhomogeneous
large deformation can at most be a superposition of
a left- and a right-moving monochromatic tachyonic mode.
A reason for this requirement is that in the Taylor series expansion of the tachyonic deformation one encounters
higher powers of tachyon operators, and the operator products must be regular.
If, on the other hand, we restrict ourselves to a small deformation, and then consider only the leading term in the
expansion,
\begin{eqnarray}\label{pert2}
Z_{def} &=& \int {\cal D}X^\mu e^{-S_0-\int \delta T} \nonumber \\
&\approx& \int {\cal D}X^\mu e^{-S_0}\left(1-\int d\vek~\delta \hat{T}_{\vek}  \int dt~e^{-i\omega_k X^0(t)+i \vek \cdot \vec{X}(t)}\right) \ ,
\end{eqnarray}
it is sufficient to consider marginal deformations and we can consider generic superpositions of different
momenta $\vek$ as long as the on-shell condition (\ref{onshell}) is satisfied.
(Here $\vek = \vek_{||}$ denotes the momentum in directions parallel to the decaying brane.)

The homogeneous rolling tachyon background $e^{X^0}$ corresponds to an exactly marginal deformation
\beq\label{roll}
 \delta S_{\rm roll} = \lambda\int dte^{X^0(t)} \
\eeq
 and leads to
the formation of pressureless tachyon matter. This was first found in \cite{Sen:2002nu}
by calculating the spacetime energy-momentum tensor
(for a slightly different tachyon profile)
in the boundary state formalism. We will base our analysis on the approach of \cite{LNT} (see also \cite{Balasubramanian:2004fz,Jokela:2005ha}), which studied
the exponential profile and calculated the energy-momentum tensor by a different approach. It is defined
as a functional derivative of the spacetime effective action with respect to the metric,
\beq
 T^{\mu\nu}(x^\mu) = \frac{-2}{\sqrt{-g}}
 \frac{\delta  S_{\rm spacetime}}{\delta g_{\mu\nu}}\Big|_{g_{\mu\nu}=\eta_{\mu\nu}} \ .
\eeq
On the other hand the action $S_{\rm spacetime}$ is given by the worldsheet disk partition function,
\beq
 S_{\rm spacetime} = Z_{\rm disk} = \int \mathcal D X^\mu e^{-S_0[g]-\delta S_{\rm roll}} \ .
\eeq
with a general space time metric in the worldsheet action,
\beq
 S_0[g]  = \frac{1}{2\pi}\int d^2z g_{\mu\nu} \partial X^\mu\bar\partial X^\nu \ .
\eeq
The energy-momentum tensor turns out to be
\beq
  T^{\mu\nu} = K ( Z_{\rm disk}'(x^0)\eta^{\mu\nu} + A^{\mu\nu}(x^0)) \ ,
\eeq
where $Z_{\rm disk}'(x^0)$ is the disk partition function (the prime indicates that the zero mode $x^\mu$ is left
unintegrated) and $A^{\mu\nu}$ is the one-point function
\beq
 A^{\mu\nu}(x^0)  =  2\left\langle :\partial X^\mu(0)\bar\partial X^\nu(0): e^{-\delta S_{\rm roll}}
 \right\rangle'
\eeq
in the rolling tachyon background (\ref{roll}). The result for the energy-momentum tensor is
\beq\label{t0}
  T_{00} = -{\cal T}_p \ ; \ T_{ij}(x^0) = \delta_{ij}{\cal T}_p (1+2\pi \lambda e^{x^0})^{-1} \ ,
\eeq
with a constant energy density and with pressure components decaying exponentially to zero at late times.


\paragraph{Perturbed energy-momentum tensor}\label{sec:deformed}
We will next calculate the spacetime energy-momentum tensor in the presence of the
initial perturbation (\ref{pert1}), (\ref{pert2}) in the rolling tachyon background (\ref{roll}).
It becomes
\beq\label{ttot}
 T_{\mu\nu}=T^{(0)}_{\mu\nu}(x^0) + \Delta T_{\mu\nu}(x) \ ,
\ee
where $T^{(0)}_{\mu\nu}(x^0)$ is the unperturbed result (\ref{t0}) and $\Delta T_{\mu\nu}(x)$ is the perturbation
which we want to calculate. It is given by
\beq\label{tpert}
 \Delta T^{\mu\nu}(x) = K \int d\vek~\delta \hat T_{\vek} \ ( \Delta Z_{\rm disk}'(x)\eta^{\mu\nu} + \Delta A^{\mu\nu}(x)) \ ,
\eeq
where the perturbation to the disk partition function $\Delta Z_{\rm disk}'$ and the perturbation $\Delta A$ involve\footnote{
Because of normal ordering, $\left\langle e^{i\vec k\cdot \vec X}\right\rangle'= e^{i\vec k\cdot \vec x}$.}
\beqa
 \Delta Z_{\rm disk}'(x) & = & \left\langle e^{\xi X^0(\tau)} e^{-\delta S_{\rm roll}}\right\rangle'
 \cdot\left\langle e^{i\vec k\cdot \vec X}\right\rangle' \\
 \Delta A^{\mu\nu}(x)        & = & 2\left\langle :\partial X^\mu(0)\bar\partial X^\nu(0):
 e^{\xi X^0(\tau)+i\vec k\cdot \vec X(\tau)} e^{-\delta S_{\rm roll}}\right\rangle' \ ,
\eeqa
where we introduced $\xi=-i\omega$. The $\Delta Z_{\rm disk}'(x)$ could be calculated from the recent
result (see Eq. (21) of \cite{Jokela:2008zh})
\beq\label{a1}
 {\cal A}_1(\xi) = \int dx^0 e^{\xi x^0}\Delta Z_{\rm disk}'(x)\Big|_{\vek=0}
  = (2\pi\lambda)^{-\xi}\Gamma(\xi)\frac{G(1+\xi)^3G(2-\xi)}{G(2\xi+1)}
\eeq
by undoing the zero mode integral. However, we will analyze both terms of (\ref{tpert}) simultaneously.
We perform a Taylor expansion of the boundary deformation. With the help of results from \cite{Jokela:2008zh}
we obtain
\beqa \label{Tmunuexp}
 \Delta T^{\mu\nu}(x)
  & = & K\int d\vek~\delta \hat T_{\vek} e^{\xi x^0+i\vec k\cdot \vec x}\sum_{N=0}^\infty (-z)^N\Bigg\{I_\xi(N)\eta^{\mu\nu} \nonumber \\\nonumber
  &   & + \frac{2}{N!}\int\frac{d\tau}{2\pi}\int\prod_{i=1}^N\frac{dt_i}{2\pi}\left\langle :\partial X^\mu(0)\bar\partial X^\nu(0): e^{\xi X'^0(\tau)+i\vec k\cdot \vec X'(\tau)} \prod_{i=1}^N e^{X'^0(t_i)} \right\rangle'   \Bigg\}\\
&\equiv& K \int d\vek~\delta \hat T_{\vek}~e^{\xi x^0+i\vec k\cdot \vec x}
\sum_{N=0}^\infty (-z)^N \left[I_\xi(N)\eta^{\mu\nu}+\Delta A_{\vek}^{\mu\nu}(N)\right]\ ,
\eeqa
where
\beqa
 I_\xi(N)
  & = & \frac{1}{N!}\int\frac{d\tau}{2\pi}\int\prod_{i=1}^N\frac{dt_i}{2\pi}\left\langle e^{\xi X'^0}\prod_{i=1}^N e^{X'^0(t_i)}\right\rangle' \\
  & = & \frac{1}{N!}\int\frac{d\tau}{2\pi}\int\prod_{i=1}^N\frac{dt_i}{2\pi}\prod_{1\leq i<j\leq N}|e^{it_i}-e^{it_j}|^2\prod_{i=1}^N|e^{it_i}-e^{i\tau}|^{2\xi} \\
  & = & \prod_{j=1}^N\frac{\Gamma(j)\Gamma(j+2\xi)}{\Gamma(j+\xi)^2} \ .
\eeqa
The more challenging task is to calculate the coefficients $\Delta A^{\mu\nu}(N)$ and then resum the
series. We start the analysis from a generating function
\beqa
 {\cal C}(\chi^{(i)}_\mu;z^{(i)},\bar z^{(i)}) &=& \left\langle :e^{\chi^{(1)}_\mu X^\mu(z^{(1)},\bar z^{(1)})} e^{\chi^{(2)}_\nu X^\nu(z^{(2)},\bar z^{(2)})}: e^{\xi X'^0(0)+i\vec k\cdot \vec X'(0)} \prod_{i=1}^N e^{X'^0(t_i)} \right\rangle' \nonumber\\
 &=& \left|1-z^{(1)}\bar z^{(2)}\right|^{\chi^{(1)}_\mu \chi^{(2)\mu}}\prod_{j=1,2}\left|1-z^{(j)}\bar z^{(j)}\right|^{\chi^{(j)}_\mu \chi^{(j)\mu}/2}  \prod_j\left|1-z^{(j)}\right|^{2\chi^{(j)}_\mu \xi^\mu}\nn\\
&& \times \prod_{i,j}\left|e^{it_i}-z^{(j)}\right|^{2\chi^{(j)0}} \prod_{i}\left|1-e^{it_i}\right|^{2\xi}\prod_{i<j}\left|e^{it_i}-e^{it_j}\right|^{2} \ ,
\eeqa
where we used rotational symmetry to fix $\tau=0$ and denoted $\xi^\mu=(\xi,-i\vec k)$.
It is then straightforward to calculate the term
\beqa
 \Delta A_{\vek}^{\mu\nu}(N) &=& \frac{2}{N!}\int\prod_{i=1}^N\frac{dt_i}{2\pi} \left.\frac{\pat^4 {\cal C}}{\pat z^{(1)} \pat \bar z^{(2)} \pat \chi^{(1)}_\mu \pat \chi^{(2)}_\nu}\right|_{z^{(j)}=\bar z^{(j)}=\chi^{(j)}=0} \ .
\eeqa
We find,
\beqa
 \Delta A_{\vek}^{00}(N) &=& \frac{2}{N!}\int\prod_{i=1}^N\frac{dt_i}{2\pi} \left(\left|\xi + \sum_i e^{it_i}\right|^2-\frac{1}{2}\right) \prod_{i}|1-e^{it_i}|^{2\xi} \prod_{i<j}|e^{it_i}-e^{it_j}|^2 \nonumber\\
           &=& \frac{N^2+2\xi N -\xi^2+2 \xi^4}{(\xi+N)^2} I_\xi(N) \\
\Delta A_{\vek}^{0j}(N)  &=& \frac{2 \xi^j}{N!}\int\prod_{i=1}^N\frac{dt_i}{2\pi} \left(\xi + \sum_i e^{it_i}\right)  \prod_{i}|1-e^{it_i}|^{2\xi} \prod_{i<j}|e^{it_i}-e^{it_j}|^2 \nonumber\\
           &=& -\frac{2 i k^j \xi^2}{\xi+N} I_\xi(N)  \\
\Delta A_{\vek}^{ij}(N)  &=& \frac{2}{N!} \left(\xi^i\xi^j+\frac{\delta^{ij}}{2}\right) \int\prod_{i=1}^N\frac{dt_i}{2\pi} \prod_{i}|1-e^{it_i}|^{2\xi} \prod_{i<j}|e^{it_i}-e^{it_j}|^2 \nonumber\\
           &=&  \left(-2 k^i k^j+\delta^{ij}\right) I_\xi(N) \ ,
\eeqa
where the integrals can be calculated by using Szeg\"o polynomials \cite{Schomerus:2008je}.\footnote{The expression for $\Delta A_{\vek}^{00}(N)$ cannot be proven directly by using the results in \cite{Schomerus:2008je}, but is found in a generalization of this calculation. We thank H. Schomerus for discussions on this point.}

In total, the series coefficients in the expansion of $\Delta T^{\mu\nu}$ are the following:
\beqa
 \Delta T^{00}(N) &=&  \frac{2\xi^2(\xi^2-1)}{(\xi+N)^2} I_\xi(N) \\
 \Delta T^{0j}(N) &=& -\frac{2 i k^j \xi^2}{N+\xi} I_\xi(N)  \\
 \Delta T^{ij}(N) &=& \left(-2 k^i k^j+ 2 \delta^{ij}\right) I_\xi(N) \ .
\eeqa
The only $N$-dependent term in $\Delta T^{ij}(N)$ is $I_\xi(N)$. Comparing with (\ref{a1}) we see that
the pressure components are proportional to ${\cal A}_1$. At this stage we can already check that the total
energy-momentum current (\ref{ttot}) is conserved. Conservation of energy requires
\beqa
 \pat_\mu T^{\mu 0}(x) &=& 2 K \int d\vek~\delta \hat T_{\vek}~\xi^2  e^{\xi x^0+i\vec k\cdot \vec x} \sum_{N=0}^\infty\frac{\xi^2-1+\vec k^2}{\xi+N} I_\xi(N)(-z)^N
\eeqa
to vanish, which is indeed the case for $\xi^2+\vec k^2=1$.
The momentum conservation equation reads
\beqa
  \pat_\mu T^{\mu i}(x) &=& -2 iK  \int d\vek~\delta \hat T_{\vek}~k^i  e^{\xi x^0+i\vec k\cdot \vec x} \sum_{N=0}^\infty \left(\xi^2+\vec k^2-1\right) I_\xi(N)(-z)^N \ ,
\eeqa
which vanishes similarly.

\paragraph{Asymptotic behavior} The energy-momentum tensor is still given in the form of a series expansion. However,
our main interest is in its asymptotic behavior as $x^0 \to \pm \infty$. In these cases we can find analytic
expressions for the leading terms. We can easily extract the leading behavior of $T^{\mu\nu}$ at
past infinity $x^0\to -\infty$. The components are given by the $N=0$ terms
\beqa
 \Delta T^{00}(x) &=& 2 K \int d\vek~\delta \hat T_{\vek}~e^{\xi x^0+i\vec k\cdot \vec x}(\xi^2-1)\left[1+{\cal O}\left(e^{x^0}\right)\right]\\
 \Delta T^{0j}(x) &=& -2 K i\int d\vek~\delta \hat T_{\vek}~ k^j e^{\xi x^0+i\vec k\cdot \vec x} \xi \left[1+{\cal O}\left(e^{x^0}\right)\right]\\
 \Delta T^{ij}(x) &=& 2 K \int d\vek~\delta \hat T_{\vek}~\left(- k^i k^j+  \delta^{ij}\right) e^{\xi x^0+i\vec k\cdot \vec x}
 \left[1+{\cal O}\left(e^{x^0}\right)\right] \ .
\eeqa
To extract the leading behavior at future infinity $x^0\to \infty$, we can use a contour integration trick
which is described in Subsection~2.2 in \cite{Jokela:2008zh}. Defining the analytic continuation of $\Delta T^{\mu\nu}(N)$ to complex values,
\be
 \Delta \tilde T^{\mu\nu}(s) \equiv \Delta T^{\mu\nu}(N \to -s)
\ee
the asymptotic behavior of $\Delta T^{\mu\nu}(x)$ in \eq{Tmunuexp} is determined by the residues of the first few poles of $ z^{-s}\Delta \tilde T^{\mu\nu}(s)/\sin(\pi s)$ on the positive real $s$-axis.
Following the analysis of \cite{Jokela:2008zh}, the residue contributions at $s=\xi,\, \xi+1$ give
\beqa
 \Delta T^{00}(x) & = &  2 K \int d\vek~\delta \hat T_{\vek}~ e^{i\vec k\cdot \vec x}(\omega_k^2+1) \omega_k^2
 \left[ x^0 {\cal A}_1(-i\omega_k) -{\cal B}(-i\omega_k)\right] +{\cal O}\left(e^{-x^0}\right)
  \nonumber\\
\Delta T^{0j}(x)  & = & -2i K \int d\vek~\delta \hat T_{\vek}~ k^j  e^{i\vec k\cdot \vec x} \omega_k^2 {\cal A}_1(-i\omega_k)\nonumber\\
                  &   & + 2i K  \int d\vek~\delta \hat T_{\vek}~\frac{ k^j  e^{i\vec k\cdot \vec x}}{(2\pi\lambda)^{-i\omega_k+1}}  \omega_k^2  \left[C(-i\omega_k)+ D(-i\omega_k) \log (2\pi\lambda) \right.
 \nonumber\\
                  &   &  \ \ \left. + x^0 D(-i\omega_k)+ D(-i\omega_k) \right] e^{-x^0}  +{\cal O}\left(e^{-2x^0}\right) \nonumber \\
 \Delta T^{ij}(x) & = & 2 K \int d\vek~\delta \hat T_{\vek}~\frac{ e^{i\vec k\cdot \vec x}}{(2\pi\lambda)^{-i\omega_k+1}}
  \left(- k^i k^j+  \delta^{ij}\right)\\ \nonumber
   & & \ \ \times \left[ C(-i\omega_k)+ D(-i\omega_k) \log (2\pi\lambda)  + x^0 D(-i\omega_k)\right]  e^{-x^0} \!
  \left[1+{\cal O}\left(e^{-x^0}\right)\right] \, .
\eeqa
First, we see that perturbations in the tachyon field give a nonhomogeneous contribution to the energy-momentum tensor of tachyon matter.
Moreover, the leading contribution to the energy density $\Delta T^{00}$ actually {\em grows linearly} as a function of time. This
is compensated by the nonzero momentum flow $\Delta T^{0i}$ that guarantees energy conservation. The pressure components $\Delta T^{ij}$
decay exponentially.
The decaying terms depend on
the coefficients 
\bea
 C(\xi) &=& \frac{\pat}{\pat s}\left[\frac{\pi}{\sin\pi s}\frac{G(\xi+1)^2}{G(2\xi+1)}\frac{\Gamma(\xi-s+2)^2G(2\xi-s+1)G(-s+1)}
  {G(\xi-s+2)^2}\right]_{s=\xi+1} \nonumber \\
 D(\xi) &=& -\frac{{\cal A}_1(\xi)}{\Gamma(-\xi)} \ ,
\eea
which were extracted from Eq.~(30) of \cite{Jokela:2008zh}. Here  ${\cal A}_1(\xi)$ is the one-point amplitude of Eq.~(\ref{a1}), and $G(\xi)$ is the Barnes $G$ function. 
The nonvanishing terms depend on the ${\cal A}_1(\xi)$, 
and on the coefficient
\beq
  {\cal B}(\xi) = \frac{\pat}{\pat s}\left[\frac{\pi(2\pi\lambda)^{-s}}{\sin\pi s}\frac{G(\xi+1)^2}{G(2\xi+1)}\frac{G(2\xi-s+1)G(-s+1)}
  {G(\xi-s+1)^2}\right]_{s=\xi} \ .
\eeq


\bigskip
\noindent

{\bf \large Acknowledgments}

We thank A. Sen for suggesting this problem to us.
N.J. has been supported in part by the Magnus Ehrnrooth Foundation and 
in part by the Israel Science Foundation under Grant No. 568/05. 
M.J. has been supported in part by the Marie Curie Excellence Grant under Contract No. MEXT-CT-2004-013510. 
E.K-V. has been supported in part by the Academy of Finland Grant No.
1127482. This work has also been supported in part by the EU 6th Framework Marie Curie Research and Training Network "UniverseNet" 
(MRTN-CT-2006-035863).


\begin{thebibliography}{99}

\bibitem{Sen:2002nu}
  A.~Sen,
  ``Rolling Tachyon,''
  JHEP {\bf 0204}, 048 (2002)
  [arXiv:hep-th/0203211].

\bibitem{Sen:2002in}
  A.~Sen,
  ``Tachyon matter,''
  JHEP {\bf 0207}, 065 (2002)
  [arXiv:hep-th/0203265].

\bibitem{Sen:2002an}
  A.~Sen,
  ``Field theory of tachyon matter,''
  Mod.\ Phys.\ Lett.\  A {\bf 17}, 1797 (2002)
  [arXiv:hep-th/0204143].

\bibitem{Frolov:2002rr}
  A.~V.~Frolov, L.~Kofman and A.~A.~Starobinsky,
  ``Prospects and problems of tachyon matter cosmology,''
  Phys.\ Lett.\  B {\bf 545}, 8 (2002)
  [arXiv:hep-th/0204187].

\bibitem{Mazumdar:2008up}
  A.~Mazumdar and H.~Stoica,
  ``Exciting gauge field and gravitons in a brane-anti-brane annihilation,''
  arXiv:0807.2570 [hep-th].

\bibitem{Recknagel:1998ih}
  A.~Recknagel and V.~Schomerus,
  ``Boundary deformation theory and moduli spaces of D-branes,''
  Nucl.\ Phys.\  B {\bf 545} (1999) 233
  [arXiv:hep-th/9811237].

\bibitem{LNT}
  F.~Larsen, A.~Naqvi and S.~Terashima,
  ``Rolling tachyons and decaying branes,''
  JHEP {\bf 0302} (2003) 039
  [arXiv:hep-th/0212248].

\bibitem{Balasubramanian:2004fz}
  V.~Balasubramanian, E.~Keski-Vakkuri, P.~Kraus and A.~Naqvi,
  ``String scattering from decaying branes,''
  Commun.\ Math.\ Phys.\  {\bf 257} (2005) 363
  [arXiv:hep-th/0404039];

\bibitem{Jokela:2005ha}
  N.~Jokela, E.~Keski-Vakkuri and J.~Majumder,
  ``On superstring disk amplitudes in a rolling tachyon background,''
  Phys.\ Rev.\  D {\bf 73} (2006) 046007
  [arXiv:hep-th/0510205].

\bibitem{Schomerus:2008je}
  H.~Schomerus,
  ``Statistical correlations in a Coulomb gas with a test charge,''
  J.\ Phys.\ A  {\bf 41} (2008) 332002
  [arXiv:0806.2602 [cond-mat.stat-mech]].


\bibitem{Jokela:2008zh}
  N.~Jokela, M.~J\"arvinen and E.~Keski-Vakkuri,
  ``N-point functions in rolling tachyon background,''
  arXiv:0806.1491 [hep-th].
\end{thebibliography}
\end{document}